\documentclass[3p,times,twocolumn]{elsarticle}

\usepackage{ecrc}
\usepackage{epsfig}

\newcommand{\beqn}{\begin{equation}}
\newcommand{\eeqn}{\end{equation}}
\newcommand{\req}[1]{Eq.\,(\ref{#1})}

\volume{00}

\firstpage{1}

\journalname{Nuclear Physics B Proceedings Supplement}

\runauth{Johann Rafelski}

\jid{nuphbp}

\jnltitlelogo{Nuclear Physics B Proceedings Supplement}

\usepackage{amsmath,amssymb}

\usepackage[figuresright]{rotating}

\begin{document}

\begin{frontmatter}




\title{Connecting QGP-Heavy Ion Physics to the Early Universe}


\author{Johann Rafelski}
\address{Department of Physics, The University of Arizona,  Tucson, 85721 USA%
}

\begin{abstract}
We discuss properties and evolution of quark-gluon plasma in the early Universe and compare to laboratory heavy ion experiments. We describe how matter and antimatter emerged from a primordial soup of quarks and gluons.   We focus our discussion on similarities and differences between the early Universe and the  laboratory experiments.
\end{abstract}
\end{frontmatter}
\vspace*{-0.5cm}
\section{Exploring quark-hadron Universe}\label{Intro}
{\small 
The standard model  of particle physics predicts that after the big-bang our Universe was   filled with  deconfined quarks, in the state of   quark-gluon plasma (QGP). This view of the early Universe is the product of over thirty years of continuous dedicated relativistic heavy ion (RHI)~\cite{Jacak:2012dx} and  lattice quantum chromo-dynamics   simulations~\cite{Fodor:2012rma}. These efforts combine and demonstrate  that in the Universe expansion QGP persisted  down to a  temperature  $T\simeq 150$\,MeV.  Considering the present day knowledge  it is not possible to imagine a different structure of the Universe before matter was formed. 

Freely propagating quarks or gluons  will never be  found in our world, yet they existed in the early Universe. To explain this  a paradigm shift has occurred in past 35 years~\cite{MKR}: we now think a change in properties of the vacuum  in the hot Universe  leads to the change  in the structure of matter. In QGP matter, particles as we know them, are dissolved into their constituents. Thus  we form in the laboratory in relativistic heavy ion collisions a small space-time domain of QGP. We confirm that quarks and gluons can roam as free particles restricted to deconfined space-time domain. We explore how matter as we know it is formed in a process called hadronization. 

The quark-Universe concepts were nascent among the group of nuclear and particle physicists who spearheaded the heavy-ion physics program in late 1970's. One of the arguments we presented was the opportunity to recreate in laboratory the deconfined form of matter, probing experimentally the physics of the early Universe and matter at extreme conditions.  The two conference proceedings of this period, Bielefeld 1980~\cite{Satz:1981tu},   and Rencontres de Moriond 1981~\cite{Jacob:1982qk} are testimony to the rapidly evolving understanding of quark matter.  This theory of quark-gluon structure of the early Universe~\cite{MKR} was born in the midst of particle and nuclear physics and took  several years to be accepted.   In the major cosmological reviews of the period such as~\cite{Dolgov:1981hv}, whenever the word `quark' appears in text, the discussion is about heavy quark remnants in the hadronic Universe, an effect which as noted above, cannot occur.

After 30 years of theoretical and experimental study of the RHI collisions, our  understanding of how matter surrounding us must have been formed progresses today from qualitative to  quantitative understanding~\cite{Letessier:2002gp,Fromerth:2002wb,Fromerth:2012fe}.  In laboratory micro-bang experiments we  study  how the QGP phase turns into a large number of hadronic particles; a detailed study of LHC experiments is presented~\cite{Petran:2013lja}. The presence of universal properties of hadronization process~\cite{Petran:2013qla} observed in several experimental conditions also clarifies  how matter was formed in the early Universe.

What we learn about hadron dynamics in the laboratory experiment clarifies how hadrons emerge and  evolve in the big-bang in QGP hadronization.  There is an even greater similarity between the micro-bang in the laboratory and the big-bang QGP stage in the early Universe.  However, the laboratory recreation of the big-bang is not exact; there are several significant differences. In this report we review the relationship between the big-bang and micro-bang with emphasis on those features which  differ. In this regard  this report  complements our other recent  related survey~\cite{Fromerth:2012fe}.%
}

\section{Quark-gluon expansion}\label{sec:expand}
When modeling the expansion of QGP for  laboratory experiments, once the particle system is thermally equilibrated, it is hard to identify entropy-generating mechanisms. This is so since particle production, e.g. flavor changing reactions, involve primarily conversion of two particles into each other at a scale where the mass does not matter, and where mass matters,  few reactions are possible. Thus QGP expansion  is an entropy-preserving process. This remark applies even more to the case of expanding Universe considering that this process is in comparison considerably slower.

The dilution of energy in expansion must thus be constrained by the conservation of entropy $S$
\beqn\label{entropy}
dE+PdV=TdS=0,\qquad dE=d(\varepsilon V).
\eeqn
As usual we will use for energy the symbol $E$, energy density $\varepsilon$,  temperature $T$, pressure $P$, volume $V$. Lower case letters denote densities, e.g. entropy density $s$, particle density $n$. From \req{entropy}  we obtain
\beqn\label{entropy1}
-\frac{d\varepsilon}{\varepsilon+P}= \frac{dV}{V}=D\frac{dR}{R}.
\eeqn
Here $R$ is the dominant size scale, and thus  $D=3$, counting the number of expanding dimensions. 

In absence of particle mass scales the particle component  (subscript `p') in energy density and pressure satisfies the relativistic equation of state and both depend on only a power of $T$
\beqn\label{EoS}
P_p(\varepsilon)=\frac{1}{3}\varepsilon_p,\quad \varepsilon_p\propto T^4
\eeqn
Combining  \req{entropy1} with \req{EoS}   produces the well-known result 
\beqn\label{EoS2}
TR=\mathrm{Const.}
\eeqn
Considering cosmic micro-wave background radiation (CMB) in current epoch \req{EoS2} conveys the remarkable constraint that while the local photon density diminishes with diminishing temperature, photons fill a larger volume and thus their number, and thus also entropy content, remains exactly preserved.  

One of tacit assumptions we made in the above discussion which lead to  \req{EoS2} is that the physical system is homogeneous.  However, for a small drop of QGP  produced in heavy ion collisions  this cannot be the case: aside from the fact that there are edges defining the size of the QGP drop, there is  energy density and local velocity field spatial dependence arising from the initial state formation process.  Thus the situation is much more complex, and the entropy preserving expansion is described in terms of  relativistic hydrodynamic flow~\cite{Huovinen:2006jp}. 

This works as follows: upon  introduction  of initial conditions which result from primary interactions of colliding partons, relativistic hydrodynamics allows to integrate the  equations towards hadronization condition.   One can learn much from this exercise, for example that it is for most part a quark-gluon fluid that flows. Even so, this  has little to do with particle interactions: it is the relativistic kinematics which connects the initial state to the designated final free-flow or hadronization condition. This is so, since  the main effect of interactions between quarks and gluons is to reduce the effective count of degrees of freedom by a factor that is rather slowly dependent on temperature~\cite{Hamieh:2000fh}, a phenomenon related to properties of quantum chromo-dynamics. The relativistic QGP fluid is a (nearly) `perfect liquid'.

Unlike heavy ion collisions, in the case of the early Universe there is at the time of QGP very likely a true  homogeneity of the Universe. On the other hand, the force of gravity controls the dynamics of expansion. This circumstance is  accounted for considering in the FRWL cosmology  the Hubble equation arising from Einstein equations:
\beqn\label{hubble}
\frac{\dot R^2}{R^2}=\frac{8\pi G}{3} \varepsilon\equiv H^2(t).
\eeqn
Combining \req{entropy1} with \req{hubble} we obtain for the evolution of the Universe  energy density 
\beqn\label{hubble1}
\dot\varepsilon^2=24\pi G \, \varepsilon (\varepsilon+P)^2
\eeqn
For purely radiative Universe we  obtain the well-known  radiative Universe power law dependence on time
\beqn\label{hubble2}
P=\frac{\varepsilon}{3} \rightarrow  \varepsilon=\frac{3}{32\pi G} \,\frac 1 {(t_0+t)^2}.
\eeqn
Integration constant $t_0$   defines the  initial energy density for  $t=0$.

Turning now to look at expansion of a bag-model like Universe~\cite{Letessier:2002gp}, we recall that we complement particle only properties by a positive volume energy ${\cal B}$ contributing a corresponding negative pressure 
\beqn\label{bag}
\varepsilon^h_p=3P^h_p\to \varepsilon^h-{\cal B}=3(P^h+{\cal B})  
\eeqn
and thus $3(P^h+\varepsilon^)h=4(\varepsilon^h-{\cal B})$, where the superscript `h' refers to hadronic component.  Note that ${\cal B}$ is just like the vacuum energy described by the Einstein term ${\cal B}\leftrightarrow \Lambda/{8\pi G}$.

Combining hadronic with non-hadronic components $\varepsilon=\varepsilon^h+\varepsilon^{\mathrm{EW}}$ we obtain the dynamical equation
\beqn\label{hubble3}
\dot\varepsilon^2=\frac{128\pi G}{3} \, \varepsilon (\varepsilon- {\cal B})^2.
\eeqn
To solve this equation we choose dimensionless variables 
\beqn\label{epsz}
\varepsilon=z^2  {\cal B}, \quad
t=x\tau_u^q, \quad \tau_u^q\equiv\sqrt{\frac{3c^2}{32\pi G {\cal B}}}.
\eeqn
The  time constant  is 
\beqn\label{hadtime}
\tau_u^q=36\mu\mathrm{s} \sqrt{\frac{{\cal B}_0}{\cal B}} 
\eeqn
where the benchmark is for ${\cal B}_0^{1/4}=195\,\mathrm{MeV}$ or equivalently $ {\cal B}_0 =0.19\,\mathrm{GeV/fm}^3$

Using \req{epsz}  we find for \req{hubble3} 
\beqn\label{hubble3b}
\frac{dz}{dx}=\pm (z^2-1),\quad z=\coth^{\pm1}(x+x_0)
\eeqn
with $x_0$ being the integration constant. A solution with decreasing energy density in time requires `+'sign 
\beqn\label{hubble4}
\varepsilon(t)={\cal B}\coth^2(x_0+t/\tau_u^q).
\eeqn
where the initial energy density at time $t=0$ is $\varepsilon(0)=\varepsilon_0={\cal B}\coth^2(x_0)$. It is common to consider $x_0\to 0$, a point-size Universe of infinite energy density. However this is a special case which will not work as we show below. Note that in the limit ${\cal B}\to 0$,  solution \req{hubble4} reduces to radiative Universe solution \req{hubble2}. 
 
We recover the usual scaling of the energy density of the Universe as follows: using in \req{entropy1} the equation of state \req{bag} and adding to it non-hadronic components, we find that 
\beqn\label{escale}
\frac{\varepsilon_0-{\cal B}}{\varepsilon-{\cal B}}=\left(\frac{R}{R_0}\right)^4
\eeqn
which reaffirms that the energy density in  particles scales with inverse 4th power of R, and thus we also recover the scaling \req{EoS2}.

In time period when the energy density is well above  $ {\cal B}$, the form of the solution \req{hubble4} is the same as the purely radiative Universe \req{hubble2}. When the energy density decreases to the scale of  $ {\cal B}$, that is on scale of $t_u$ time $t$ is not small, the energy density of the Universe is nearly constant and dominated by the vacuum energy ${\cal B}$. At that condition with a proper choice for ${\cal B}$ the transition to hadron Universe must occur. The  time constant for this expansion towards transition is as stated in \req{hadtime}. 

The meaning of this time scale $\tau_u^q$ \req{hadtime} is understood recalling that e.g. at  a temperature scale a 1000 times larger than the hadronic transition, $T_{\rm EW}\simeq 150$\,GeV  the electroweak phase was symmetric - and all quarks were massless. In order to describe a reduction of temperature by a factor 1000 and thus energy density by 12 orders of magnitude, the initial time would need to be according to the scaling we discussed
\beqn\label{Size2}
\tau_u^\mathrm{EW}=\frac{\tau_u^q}{10^6},
\eeqn
thus the electro-weak transition time scale is measured in terms of 10's of picoseconds. Seen this we realize that as we go from pico to micro seconds we pass through the entire QGP era of the Universe.  $\tau_u^q$ is thus the effective lifespan of the quark Universe.

\section{Baryon asymmetry}\label{sec:baryon}
We are interested in comparing  the Universe entropy content per baryon $s/n_B=S/N_B$ with what is achieved in laboratory experiments. Assuming that the baryon number is  conserved   and recalling that the expansion of the Universe is adiabatic, the ratio $s/n_B$, a dimensionless number, is preserved in the Universe evolution.

Considering particle components in the Universe when  the entropy is dominated  by  photons, $e,\mu,\tau$-neutrinos, and electron-positron pairs, which is the case e.g in the temperature range   $50> T> 2$ MeV, we have
\beqn\label{SperB1}
\frac{s}{n_B}=\frac{1}{n_B}\sum_{i=\gamma,\nu,e}s_i
=\frac{n_\gamma}{n_B},\left(\frac{s_{\gamma}}{n_{\gamma}}+\frac{n_\nu}{n_\gamma}\frac{s_\nu}{n_\nu}+\frac{n_e}{n_\gamma}\frac{s_e}{n_e}\right) 
\eeqn
For photons $(s/n)_{\rm boson}=3.60$ and for neutrinos as well  in the above temperature domain where electrons are effectively massless, $(s/n)_{\rm fermion}=4.20$.  For relative quantum distribution massless particle densities  we recall the Riemann function relation  $n_{\rm fermion}/n_{\rm boson} =\eta(3)/\zeta(3)=3/4$.  Canceling the spin factor 2 between photons and fermions, we are left with the count of  the number of fermions, anti-fermions, but  only  single-handed neutrinos of the three flavor. We find
\beqn\label{SperBcalc}
\frac{s}{n_B}\frac{n_B}{n_\gamma} = 3.60 + 3\frac{3}{4} 4.20+2\frac{3}{4} 4.20=19.35.
\eeqn 
 
The value of ${n_B}/{n_\gamma}$ can be obtained considering BBN nucleo-synthesis yields in the BBN era, and later, we can also obtain it at the recombination period. We adopt the value
\beqn\label{Bpergamma}
\eta\equiv \frac{B}{n_\gamma}=6.27\pm0.34\times 10^{-10}, 
\eeqn 
see e.g. figure 4 of Ref.~\cite{Steigman:2012ve}, where in first approximation we assume that all entropy in the annihilating electron-positron pairs reheats  photons. Using \req{Bpergamma} we obtain 
\beqn\label{soverb}
\frac{s}{n_B}\equiv\frac{S}{N_B}=\frac{19.35}{\eta}\simeq 3\times 10^{10}\,.
\eeqn 
Possible reheating of neutrinos in $e^+e^-$-annihilation prior to BBN introduces  some uncertainty here, and is a topic which remains under current investigation~\cite{Steigman:2012ve,Birrell:2013gpa}.

Turning attention now to the heavy ion collisions we note that it is understood that as the energy of the collision  increases, the number of quarks retained in the central rapidity i.e. CM frame of the collision domain diminishes. One can think of this situation akin to a shot through a thin target: there is energy deposition; however, since the target is thin, when the collision occurs at a sufficiently high energy  these parton-bullets cannot be stopped, and thus the baryon number they carry was expected to leave the interaction region. One expects that heavy ion collisions are producing a QGP that is as   free of net baryon number as is  the early Universe. The observation of a significant stopping of baryon number in collisions of 100+100 GeV per nucleon for heaviest nuclei at RHIC remains in need of a convincing explanation.

At LHC where experiments at 1380+1380 GeV per nucleon were recently carried out, the baryon yield in the central collision region is so small that its measurement has   become an experimental challenge.  A recent in-depth analysis produces as an  estimate~\cite{Petran:2013lja}  
\beqn\label{soverbHI}
\left.\frac{s}{n_B}\right|_{\rm lab}\simeq 5000. 
\eeqn 
The QGP we make today at LHC   is by 6 order of magnitude baryon richer. By considering experiments at a lower energy we can achieve QGP with significantly higher baryon content.  For some physics challenges this could  be an advantage:  in the big-bang the large volume available  allows  the smallest asymmetry to be amplified. In the lab the smallness of the available volume of QGP requires a much greater baryon asymmetry for an effect of interest to become visible.

\section{Size of hadronizing  QGP}\label{sec:size}
But how big was the Universe at the time of hadronization? To answer this we consider how much energy and mass is in the quark Universe that  hadronizes.  Consider first  a Universe which was point-like and expanded radiatively to  the size given by $R_u^q=c\tau_u^q\simeq 10$\,km \req{hadtime}.  At hadronization, the energy density scale is provided by $4{\cal B}$. Multiplying with $4\pi(R_u^q)^3/3$  we obtain the mass equivalent of $10^{57}$--$10^{58}$ protons. Most of this stuff, down to a fraction $10^{-10}$, annihilates and feeds Universe expansion and present day CMB. The insight we gain is that even if all this hadronization energy were to go to make matter we see out the window, we still would be far from being able to describe all matter. We are missing  more than 40  orders of magnitude. The Universe at the time of QGP hadronization must have been very much larger which means that before the QGP era it was already  quite large and not a  `point'.

The question --- how big is the hadronizing Universe which provides the visible matter surrounding us? --- leads us back to consider the  entropy content in the Universe.  We recall that we know  the entropy per baryon $s/n_B$. Moreover, the PDG~\cite{PDG12} suggests that today we have $n_B\simeq 0.25\mathrm{m}^{-3}$ in the Universe. Thus the present day entropy density is 
\beqn\label{spresent} 
s=\frac{s}{n_B} n_B=0.75\times 10^{10}\mathrm{m}^{-3} \equiv \frac{s_h}{(z_h+1)^3}. 
\eeqn
In the last relation  we connected by the  redshift factor $z_h$   to entropy content at  time of hadronization. Note that in adiabatic expansion of the Universe the scaling with $z$ of entropy density must be just like that of a conserved particle number density. That this is true one can easily see contemplating how the ratio ${s}/{n_B}$ can remain unchanged during the evolution   history of the Universe.

The entropy density at hadronization $s_h$ can be seen as composed of two components: 1) $\sigma_h^q$ is  the strong interaction part  originating in quarks and gluons.  Due to the color degree of freedom it is the dominant component, over the second component; 2)  $\sigma_h^{\mathrm{EW}}$, which includes contribution of effectively non-interacting electro-weak degrees of freedom which are: photons, electrons, muons, and the three single-handed neutrinos. The quark-gluon component  is measured both in lattice gauge theory studies~\cite{Fodor:2012rma} and in hadronization studies of QGP~\cite{Petran:2013lja,Petran:2013qla}, where  
\beqn\label{shhad} 
s_h^q \simeq \frac{3.5}{\mathrm{fm}^3}.
\eeqn
Counting the free particle degrees of freedom one finds that EW components   contribute  $\sigma_h^{\mathrm{EW}}/s_h^q\simeq 0.3$. However, actual contribution is  at about 40\% -- 50\%, where the uncertainty comes from the exact magnitude of the strong interaction part of entropy density which is not fully described   counting degrees of freedom and which changes rapidly at hadronization condition. We will here use 
\beqn\label{stoth}
s_h =s_h^q+s_h^{\mathrm{EW}}\simeq   \frac{5}{\mathrm{fm}^3}\,.
\eeqn

Using \req{stoth} in \req{spresent} we obtain 
\beqn\label{zhad} 
z_h=0.9\times  10^{12}
\eeqn
This value checks out with a simple qualitative estimate:  the radiative expansion dominated portion, $\simeq 10^9$   connects the recombination era $T_r=0.25$\,eV with the hadronization temperature $T_h=150$\,MeV,  and there is remaining factor  $z_r\simeq 1000$ connecting the present to recombination.
Among the near future tasks in study of QGP Universe is a more precise determination of $z_h$.  

Given the entropy density at hadronization \req{stoth} and entropy per baryon \req{soverb} (or simply the scaling with $z_h$ of baryon density) we  also know that at the time of hadronization the net baryon density of the Universe was relatively small
\beqn\label{bhadroniz} 
n_B^h\simeq (z_h+1)^3  n_B\simeq \frac 2 3 10^{36}\frac 1 4 \mathrm{m}^{-3}=1.67\times 10^{-10}\mathrm{fm}^{-3}.
\eeqn
In order to get   a  sufficiently large number of baryons  we must multiply \req{bhadroniz} by a rather gigantic volume!  For example if we wish to have a baryon content $N_B= 10^{80}$  the Universe volume at time of hadronization was 
\beqn\label{VUniverse2}
V_u=\frac 3 5 \times  10^{90} \mathrm{fm}^3, 
\eeqn
which provides us with a radius $R_u$ and time scale $\tau_u$,
\beqn\label{VUniverse3}
R_u\simeq 0.5\times 10^{30}\mathrm{fm},\quad \tau_u=R_u/c=1.5\times 10^{6}\mathrm{s}.
\eeqn
We are not taking sides here in any ongoing discussion regarding the baryon inventory in the Universe. Irrespective of the precise baryon inventory value, the size of the quark Universe at the time of hadronization is large: the light takes a month  to cross the diameter of the hadronizing QGP domain.

In comparison to the QGP Universe which as we have now shown must be very large, the laboratory micro-bang is truly small.  The laboratory micro-bang scale is governed by nuclear size and a more precise evaluation which depends a bit on collision energy and geometry condition does not change the fact that the micro-bang radius is 29 orders of magnitude smaller compared to \req{VUniverse3}.

\section{Antimatter annihilation}\label{sec:QGPHG}
\subsection{Evolution constraints} \label{ssec:cons} 
In the early Universe when the hadron matter phase emerges from the QGP we have both matter and antimatter present present in large abundance, and as the Universe expands and the ambient temperature drops,  for quite some time there is  annihilation between matter and antimatter.  A more detailed study of this era is presented in Ref.\cite{Fromerth:2012fe}, we will develop two important physics points below, the evolution of hadron abundances and the potential for abundance distillation. 

The situation is very different in the micro-bang. Given the relatively small volume of laboratory QGP, one of the interesting outcomes is that after hadronization particles begin to free-stream very early which prevents antimatter annihilation and  we produce antimatter abundantly. For further details we refer the reader to the recent analysis of the LHC hadron production results in Refs.\cite{Petran:2013lja,Petran:2013qla}. However, this also means that big-bang  physics phenomena which depend on a slow distillation process will not be easily visible in the micro-bang model.

The particle composition in QGP, both,  in the hadronic Universe  as well as  in the micro-bang must satisfy three global constraints \\
1) {\it Charge neutrality} ($Q = 0$) 
\beqn\label{Q0}
n_Q\equiv \sum_f\, Q_f\, n_f (\mu_f, T)=0, 
\eeqn
where $Q_i$ is the charge of species $f$, and the sum is over all particle species present in the considered particle phase.\\
2) {\it Net lepton number equals net baryon number} ($L = B$) is phenomenologically motivated in the context of baryo-genesis.  This leads to the constraint
\begin{equation}\label{LB0}
n_L - n_B\equiv \sum_f\, (L_f - B_f)\, n_f (\mu_f, T)=0 ,
\end{equation}
where $L_f$ and $B_f$ are the lepton and baryon numbers of species $f$. This condition is of course not proved by experiment and future studies must consider drastically different variants.\\
3) {\it A given value of entropy-per-baryon} ($S/B$). 
The value of $s/n_B$ is obtained from ${n_\gamma}/{n_B}$ as already discussed in section \ref{sec:baryon}.

Considering how many particles can be present we need to find additional constraints. These arise from reactions between different components. For any reaction $\nu_f A_f = 0$, where $\nu_f$ are the reaction equation coefficients of the chemical species $A_f$, chemical equilibrium occurs when $\nu_f \mu_f = 0$, which follows from minimizing the Gibbs free energy. Here $\mu_f$ are chemical potentials which we now consider.

\subsection{Chemical potentials} \label{ssec:evol} 
In a system of non-interacting particles, the chemical potential $\mu_f$ of each species $f$ is independent of the chemical potentials of other species, resulting in a large number of free parameters.  Reactions between components result in   reaction equilibrium.  In the early Universe and in particular in the hadronic epoch following on the QGP hadronization three different equilibrium conditions are recognized given differences in pertinent reaction cross sections:\\
1) \underline{The kinetic equilibrium} (also called thermal equilibrium): particles equipartition energy by means of collisions that do not change particle number or undergo changes akin to chemical reactions. This type of interaction assures that particles have the Bose or Fermi (Boltzmann) momentum distribution. Their number can differ from naive expectation.\\ 
2) \underline{Relative chemical equilibrium}  is specific to hadron world: individual quarks can flow between hadrons but quark pairs are not produced. In relative chemical equilibrium, the chemical potential of hadrons is equal to the sum of the chemical potentials of their constituent quarks.  For example, $\Sigma^0 (uds)$ has chemical potential $\mu_{\Sigma^0}=\mu_u + \mu_d + \mu_s=3\,\mu_d - \Delta \mu_l$.\\
3) (Absolute) \underline{chemical equilibrium}:  quark-antiquark pair production processes are fast enough to assure that the yields of  hadrons are in equilibrium. This equilibrium also addresses the weak or electromagnetic  interaction process that connects  hadrons with leptons. When pairs are in equilibrium, i.e., the reaction $f + \bar{f} \rightleftharpoons 2 \gamma$ proceeds freely in both directions.  Therefore, $\mu_f = -\mu_{\bar{f}}$ whenever chemical and thermal equilibrium is attained.  

It turns out that only weak interactions  are  weak enough in the early Universe to freeze-out and this occurs relatively late, beyond our current interest.  When the system is chemically equilibrated with respect to weak interactions~\cite{Glendenning:2000wn}:
\beqn
\mu_u=\mu_d - \Delta \mu_l,\quad \mu_s=\mu_d, \quad \mu_B\equiv\frac{3}{2}(\mu_d +\mu_u)
\eeqn
where
\beqn
\Delta \mu_l=\mu_i - \mu_{\nu_i},\quad i=e,\mu,\tau\,.
\eeqn
and neutrino oscillations   imply that neutrino number is exchanged between flavors $\nu_e \rightleftharpoons \nu_\mu \rightleftharpoons \nu_\tau$ and hence
\beqn
\mu_{\nu_e} = \mu_{\nu_{\mu}} =\mu_{\nu_{\tau}} \equiv \mu_\nu.
\eeqn

The strong and electromagnetic interactions remain always faster than the expansion of the Universe.  The situation is grossly different in the micro-bang. The weak and electromagnetic interactions are entirely decoupled from the micro-bang. The expansion of QGP is fast enough to worry about detailed study of hadron reaction decoupling and chemical non-equilibrium among hadrons.


Reactions between particles expressed in terms of equilibrium chemical conditions supplemented by conservation conditions always close the system of chemical rate equations. Thus one can find a unique solution and obtain all particle abundances present. For the early Universe these equations were for the first time established and solved numerically in Ref.\cite{Fromerth:2002wb}. Some of these results were presented for the first time in recent survey   Ref.\cite{Fromerth:2012fe}.  We recall a few important results:  Strangeness, present in kaons, persists down to $T=10~{\rm MeV}$.  Pion and muon evolve in chemical equilibrium as we briefly discuss in next section \ref{sec:pionlepton}, their density falling below density of nucleons only below $T\simeq 6$~MeV. 

\begin{figure}
  \centerline{\includegraphics[width=0.45\textwidth]{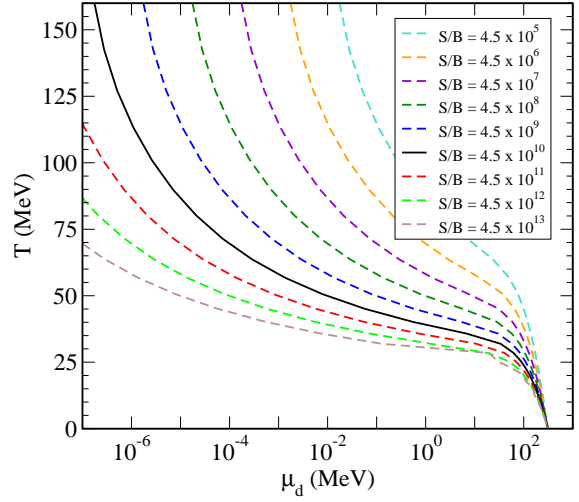}}
\caption{The chemical potential  $\mu_d $ evolves towards common value $\mu_d\simeq m/3$ for any value of $S/b$~\cite{FromerthUnp}.\label{chem_pot} }
\end{figure}

In order to describe the low net baryon density \req{bhadroniz} the value of the baryon chemical potential just before the phase transition is a tiny $10^{-10}$ fraction of the temperature
\beqn\label{preHGmuB}
\mu_B = 0.33^{+0.11}_{-0.08}~{\rm eV}.
\eeqn
As temperature decreases  $\mu_d$ approaches (weighted) one-third the nucleon mass $(2m_n-m_p)/3=313.6~{\rm MeV}$ reflecting the dominance of protons and neutrons in their classical Boltzmann limit $T\ll m_f$, see Fig.~\ref{chem_pot}. The different lines step each by an order of magnitude the value of entropy per baryon \req{Bpergamma}, and we see a significant variance in behavior. However,   such a large range of $S/B$ requires a considerable change in baryon or entropy content prior to BBN which occurs at the final convergence point in Fig.~\ref{chem_pot}.

\subsection{QGP--HG transformation and distillation}\label{MixedPhase}
The conversion of QGP to hadrons can  proceed by formation of  separate  domains of both phases. The issue is that in HG the entropy density is lower than it is in QGP and the transition from QGP to HG in chemical equilibrium can only occur if in conversion from QGP to HG   the volume occupied by HG is much larger. Thus the total volume will grow while the ambient temperature remains practically unchanged. While this is occurring distillation of the quantities we needed to conserve and in particular the  baryon and  charge density  can  occur. 

Considering that  matter-antimatter symmetry has been very slightly broken, as a function of $f_{\rm HG}(t)$  there will in general be further asymmetry developing e.g. in the baryon distribution, which is different in two phases for same statistical parameters. To date nobody has carried out a realistic study of the dynamical Universe evolution allowing for proper distillation of the physical properties between the phases, along with annihilation of the high initial matter antimatter content. This remains an interesting research task for the  near future. 

A first insight~\cite{Fromerth:2002wb,Fromerth:2012fe} is obtained considering the total partition function  parametrized as
\beqn
\ln{Z_{\rm tot}}=f_{\rm HG}\, 
\ln{Z_{\rm HG}}+(1 - f_{\rm HG})\, \ln{Z_{\rm QGP}} ,
\eeqn
in which $f_{\rm HG}(t)$ represents the fraction of total volume occupied by the HG phase. In a model we can assume that $f_{\rm HG}$ evolves linearly in time and that the total duration of the phase transformation is e.g. $10~\mu{\rm s}$, the actual dynamics of the transition is beyond the current discussion scope. Note that  \req{Q0} is now generalized as
\begin{align}\label{Qmixed}
Q = 0 & = n_Q^{\rm QGP}\, V_{\rm QGP}+n_Q^{\rm HG}\, V_{\rm HG}, \\
  &=  V_{\rm tot} \left[ (1-f_{\rm HG})\, n_Q^{\rm QGP}+f_{\rm HG}\, n_Q^{\rm HG} \right]. \nonumber
\end{align}
and analogous expressions hold for Eqs.\,\eqref{LB0} applies.  For each $f_{\rm HG}$ the net charge per baryon $n_Q/n_B$ is calculated in each phase as a function $f_{\rm HG}$, which is independent of the additional assumptions.  Protons and neutrons being the lowest excitations in the HG phase, the HG takes on a positive charge as soon as the transformation begins.  The QGP therefore takes on a negative charge density, which is initially tiny since it occupies the larger volume, yet it can cause large variation in local electric potentials,  a point needing much future attention.

\section{Pion and lepton equilibration: $50 \gtrsim T \gtrsim 2~{\rm MeV}$}
\label{sec:pionlepton}
It is important to realize that hadrons always are a part of the evolving Universe, a point not much discussed. The reaction allowing  presence of hadrons in  a `cold' evolving Universe is~\cite{Kuznetsova:2008jt}
\beqn\label{pi02gamma}
\pi^0\rightleftharpoons \gamma+\gamma,
\eeqn
Comparing relaxation time for this reaction $\tau_\pi$ to $\tau_u \equiv 1/H$  shows that $\pi^0$ remains in chemical equilibrium even as its thermal number density gradually decreases, consistently with  falling thermal production rates. This phenomenon can be attributed to the high population of photons, within which it remains probable to find photons of high enough energy to fuse into $\pi^0$, and as the number of high energy photons decreases, described by photons' Planck distribution, so does the number of $\pi_0$ which need to be maintained.  

Two-to-two reactions maintain equilibrium within and between charged pion populations~\cite{Kuznetsova:2010pi}
\beqn\label{chargedpipi0}
\pi^0+\pi^0 \rightleftharpoons \pi^++\pi^-, \quad 
\gamma+\gamma \rightleftharpoons \pi^++\pi^-
\eeqn
and lepton populations
\beqn\label{2lto2l}
\begin{split}
e^++e^-\rightleftharpoons \mu^++\mu^-,
\quad \gamma+\gamma\rightleftharpoons l^++l^-,\\
\pi^++\pi^-\rightleftharpoons l^++l^-,~~(l=\mu,e).
\end{split}
\eeqn
The chemical relaxation times of each of these reactions are faster than the expansion of the Universe. 

\section{Conclusions}\label{sec:concl}
Today we can look back in the history of the Universe by observing free-streaming photons  created at $T_r=0.26$\,eV (about 3000 K) when electrons and nuclei recombined. The Universe was about 1000 times smaller in size. The next snapshot comes from study of the abundances of light isotopes made in the period of the big-bang nucleosynthesis   when temperatures are another factor 10,000 times greater, at $T=30$\,keV and more. Looking beyond this we can not `see' anything until we get to the QGP, recreated in  the  micro-bang, corresponding to an era when the Universe was  10,000 hotter and smaller in size:  the QGP hadronized at $T\simeq 150$\,MeV at $z_h=0.9\times 10^{12}$, see  section \ref{sec:size} and \req{zhad}. The big-bang QGP is tiny: in size scale we are missing 29 orders of magnitude.  Even if we assume a point-like initial condition for the big-bang QGP, the size of the Universe would be 10km~\cite{Letessier:2002gp} and the scale of big-bang still differs by 18 orders of magnitude from the micro-bang.

It is a big question if from the tiny micro-bang laboratory model of the early Universe we can  learn everything we need to know about the  big-bang QGP and hadron phases.  In the  QGP micro-bang we only see   a tiny piece of the   Universe and as our discussion shows there are many differences to overcome, before we can connect with the big-bang. However,  we create QGP  in the laboratory and  we can take time and effort to understand and generalize what we learned, so as to be able to use it to understand the hadron Universe  once it emerges from the EW transition beginning e.g. at 0.1ns passing through the QGP hadronization near $30\mu$s and reaching BBN at a few seconds of age.

We have discussed  that hadrons remain in chemical equilibrium throughout the evolution of the Universe. The same mechanism that allows $\pi_0$ to rapidly decay in reaction \req{pi02gamma} is  present when two photons collide in the thermally equilibrated Universe, and photon `fusion' reactions fill any missing $\pi_0$~\cite{Kuznetsova:2008jt,Kuznetsova:2010pi}. Given that hadron chemical equilibrium is maintained, one can use methods of hadro-chemistry to study particle abundances. We have shown how to compute the values of the quark and lepton chemical potentials that yield the observed matter--antimatter asymmetry in the early Universe after hadronization and equilibration of the HG. 

The non-zero chemical potentials drive charge distillation during the phase transformation, with the QGP and HG having negative and positive charge densities, respectively. Large Coulomb fields may be present across   phase boundaries in case electrons and positrons must be part of the neutralization process, these very  low mass particles cannot follow the sharp hadronic phase boundaries precisely. Appearance of strong Coulomb fields  can play a significant role in amplifying pre-existent baryon asymmetry, thus a much smaller net baryon asymmetry could be amplified to values we see, a point that certainly needs much further study.

To finish, we note there three pillars of the field of QGP: the creation, observation, exploration of QGP, a very dense new phase of matter; the study of  confinement of quarks; and the creation of matter from energy, that is hadronization. These   topics as we have discussed here connect directly the experimental relativistic heavy ion program with the physics of the early Universe. The study of QGP in the laboratory  is today  the only experimentally accessible approach  to improve the understanding of the early Universe at a temperature that is a billion times greater than the ion-electron recombination condition   at $T_r\simeq 0.25$\,eV.

\vskip4mm
\noindent {\it Acknowledgments}  I thank J. Birrell for discussions  about size of the Universe and point-like initial conditions  considered in~\cite{Letessier:2002gp}.
This work has been supported by a grant from the U.S. Department of Energy, DE-FG02-04ER41318.

\end{document}